\documentclass[12pt]{iopart}

\usepackage{graphicx,bm}
\usepackage{txfonts}
\usepackage{dsfont}
\usepackage{color}
\usepackage{hyperref}

\newcommand{\rp}[1]{(\ref{#1})}

\newcommand{\abs}[1]{\left|{#1}\right|}

\newcommand{\av}[1]{\left\langle #1 \right\rangle}

\newcommand{\al}[1]{^{(#1)}}
\newcommand{\da}{^\dagger}

\newcommand{\ppt}[1]{\left( #1 \right)}
\newcommand{\pq}[1]{\left[ #1 \right]}
\newcommand{\pg}[1]{\left\{ #1 \right\}}

\newcommand{\lpq}[1]{\left[ #1 \right.}
\newcommand{\lpg}[1]{\left\{ #1 \right.}

\newcommand{\rpq}[1]{\left. #1 \right]}
\newcommand{\rpg}[1]{\left. #1 \right\}}
\newcommand{\ee}{{\rm e}}
\newcommand{\ii}{{\rm i}}
\newcommand{\dd}{{\rm d}}

\usepackage{bbold}

\newcommand{\nn}{{\nonumber}}

\newcommand{\mmat}[2]{
                      \begin{array}{#1}
                       #2
                       \end{array}  }

\newcommand{\BBB}{{\cal B}}

\newcommand{\FF}{{\cal F}}

\usepackage[normalem]{ulem}

\newcommand{\stkout}[1]{\ifmmode\textrm{\sout{\ensuremath{#1}}}\else\sout{#1}\fi}

\begin{document}



\title{
Feedback-enabled Microwave Quantum Illumination\\
}

\author{Mehri Sadat Ebrahimi$^{1,2}$, Stefano Zippilli$^1$, David Vitali$^{1,3,4}$}
\address{$^1$ School of Science and Technology, Physics Division, University of Camerino, I-62032 Camerino (MC), Italy\\
$^2$ Department of Physics, University of Isfahan, Hezar-Jerib, 81746-73441, Isfahan, Iran \\
$^3$ INFN, Sezione di Perugia, Italy\\
$^{4}$ CNR-INO, L.go Enrico Fermi 6, I-50125 Firenze, Italy}


\vspace{10pt}
\begin{indented}
\item[]\today
\end{indented}

\begin{abstract}
A simple feedback scheme can be used to operate efficiently a microwave-quantum-illumination device based on electro-optomechanical systems also in regimes in which excess dissipation would, otherwise, prevent to outperform the optimal classical illumination protocol with the same transmitted energy.
\end{abstract}

%
%
%
%
%

\section{Introduction}

Quantum illumination (QI) is a powerful quantum-optical sensing technique which, unlike other quantum optical techniques, is particularly resilient to background noise~\cite{Lloyd2008,QI with Gaussian states,pirandola2018,shapiro2020,sorelli2021,Torrome2021}.
The main purpose of QI is to detect a low-reflectivity object in the presence of very bright thermal noise.
It makes use of two entangled modes of the electromagnetic field. One, the signal, is sent to probe the target region, while the other, the idler, is retained at the source. Finally a joint measurement of the idler and of the field possibly reflected by the target is used to determine the presence of the target itself.
Although the entanglement is destroyed in the round trip from the target, the remaining signal-idler correlations are still sufficient to outperform any classical illumination system of the same transmitted energy~\cite{QI with Gaussian states, Gaussian state QI Guha,Discord for QI,Karsa1,Karsa2}.

The main feature of quantum illumination is that it performs better in the presence of intense thermal radiation (noise). Even if several studies have been focused on the optical regime~\cite{Optical QI 1, Optical QI 2, Optical QI 3} where the number of thermal photons is negligible,
protocols operating in the microwave regime~\cite{Microwave QI 1, Balaji1,Balaji2,Balaji3,Microwave QI 2,Karsa-mw} have attracted particular attention because of the naturally occurring bright microwave background radiation.

In the first theoretical proposal of microwave QI~\cite{Microwave QI 1}, an electro-optomechanical (EOM) system with single-sided cavities has been used as a transmitter, which generates the entangled signal and idler. The signal is a microwave field used to probe the target, while the idler is optical~\cite{Barzanjeh2012}.
In general QI works better with high-efficiency photodetectors which up to now are available mostly for the optical domain only. For this reason,
in this protocol, another electro-optomechanical system, equal to the first one, is used to up-convert the reflected microwave signal into an optical field and the final joint measurement is done, in the optical regime, between the optical idler and the converted signal.

In this paper, we consider the same scheme as in Ref.~\cite{Microwave QI 1} with two
EOM devices as a transmitter and a receiver. Differently from the old proposal, here we consider the more realistic case of two-sided cavities. This is a very natural choice realized, for example, in the case of Fabry-Perot cavities. In this case, the added noise due to the additional decay channels, suppresses the performance of the QI protocol, which may be no more able to outperform the classical illumination benchmark given by homodyne detection of a classical coherent field~\cite{QI with Gaussian states, Gaussian state QI Guha}. To compensate for this suppression we propose to add a feedback-loop from the optical to the microwave cavity of the EOM transmitter, which recovers part of the correlations, otherwise lost in the unused cavity decay channel. This approach is inspired by other studies on cavity optomechanics within a feedback-loop which have demonstrated the efficiency of similar setups to control the system dynamics~\cite{feedback-loop sample 1, feedback-loop sample 2, feedback-loop sample 3, feedback-loop sample 4, feedback-loop sample 5, feedback-loop sample 6}. We show that by properly adjusting the feedback-loop and appropriately selecting the system parameters, our proposed scheme can efficiently outperform any classical system of the same transmitted energy.

In Sec.~\ref{system} we provide the model details and the basic equations, while in Sec.~\ref{Sec.MQI} we describe the QI protocol and derive the main relevant quantities. In Sec.~\ref{num} we provide all the numerical results, and in Sec.~\ref{concl} we provide some concluding remarks.

\section{The system}\label{system}

\begin{figure}\centering
\includegraphics[width=15cm]{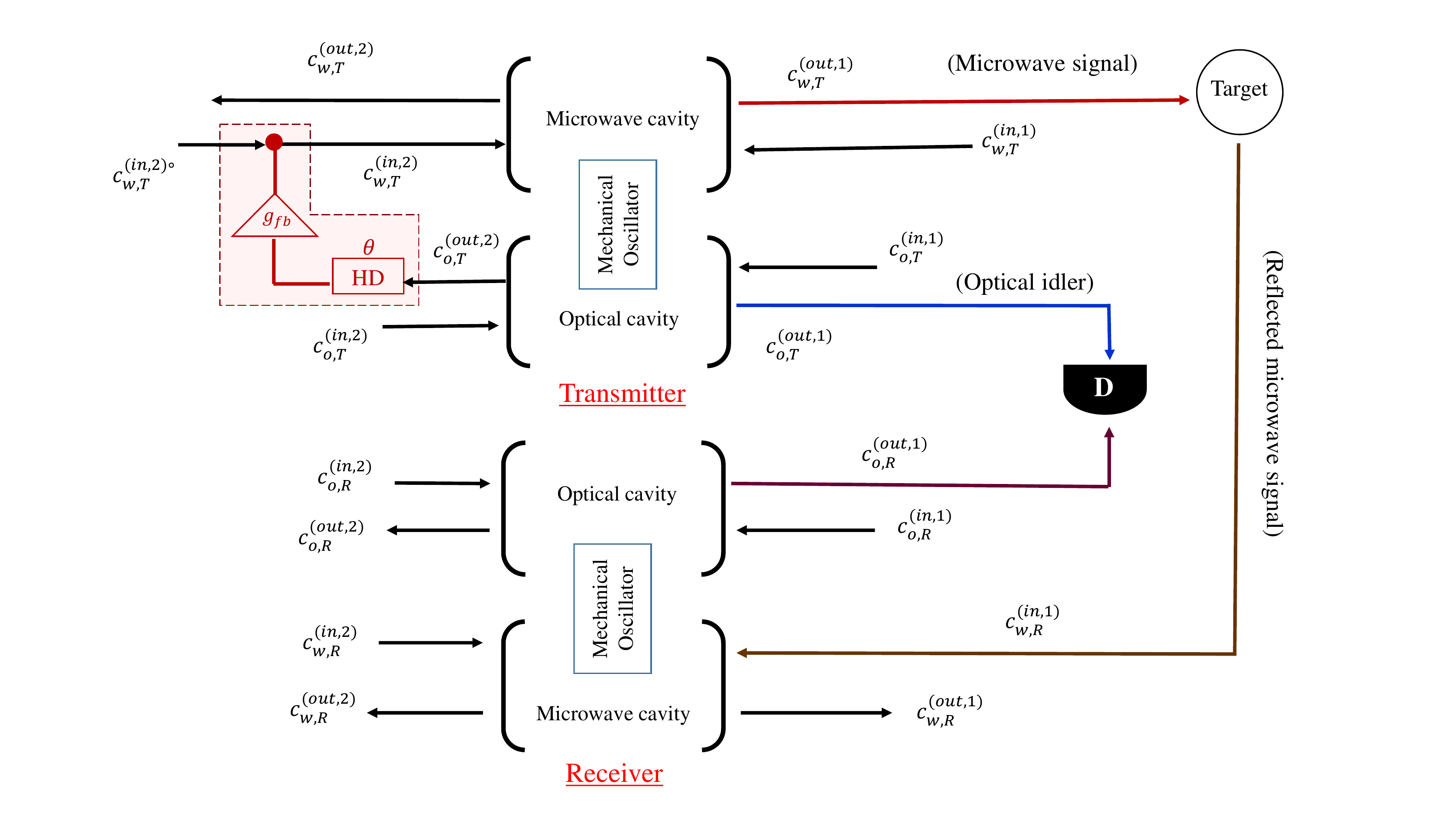}
\caption{(Color online)
Schematic of the proposed microwave QI system.
Enclosed by the dashed line in the upper left part, the feedback setup where the second optical output of the 
transmitter
is measured by homodyne detection (HD) with homodyne phase $\theta$
(used to control the feedback phase $\phi$) and the corresponding signal is used to modulate the amplitude of the second microwave input with feedback gain $g_{fb}$.
The rest of the device is similar to the one studied in Ref.~\cite{Microwave QI 1}: the microwave signal illuminates the target; The optical idler and the signal converted and phase conjugated by the receiver are combined and detected (D) to discriminate the presence of the target.
}
\label{Fig1}
\end{figure}

We consider two similar EOM systems which play the role of the transmitter and of the receiver (in the rest of the paper we will use the symbols $T$ and $R$ to distinguish between quantities pertaining to one and the other), see Fig.~\ref{Fig1}. In each EOM system, a mechanical resonator with frequency $\omega_{M,\xi}$ and dissipation rate $\gamma_{M,\xi}$ (with $\xi\in\pg{T,R}$), interacts with a microwave mode of a superconducting cavity at frequency $\omega_{w,\xi}$ and with an optical mode of a Fabry-Perot cavity at frequency $\omega_{o,\xi}$.
We assume that the EOM systems are in a cryogenic environment in order to minimize the number of thermal microwave photons.
The two modes lose photons at rate $\kappa_{i,\xi}$ (with $i\in\pg{w,o}$ used to distinguish between microwave and optical modes), and are driven
by two coherent fields at frequencies $\omega_{i,\xi}\al{d}$, detuned by $\Delta_{i,\xi}\equiv \omega_{i,\xi}-\omega_{i,\xi}\al{d}$ from the two cavities.
We assume sufficiently large driving powers, so that we can employ the standard linearized description for the optomechanical interaction. Specifically, we describe the microwave, optical and mechanical modes, respectively, in terms of the bosonic annihilation operators $\hat{c}_{w,\xi}$, $\hat{c}_{o,\xi}$ and $\hat{b}_\xi$ for the fluctuations around the corresponding average fields.
These operators fulfill the following quantum Langevin equations (QLEs)
\begin{eqnarray}\label{QLE 1w}
\dot{\hat{c}}_{w,\xi}(t)&=&
-\ppt{
\kappa_{w,\xi}+i\Delta_{w,\xi}
}\ \hat{c}_{w,\xi}
-iG_{w,\xi}\ \ppt{
\hat{b}_\xi+\hat{b}_\xi^{\dagger}
}+\sqrt{2\kappa_{w,\xi}}\  \hat{c}_{w,\xi}^{(in)}\ ,
\\\label{QLE 1o}
\dot{\hat{c}}_{o,\xi}(t)&=&-\ppt{
\kappa_{o,\xi}+i\Delta_{o,\xi}
}\ \hat{c}_{o,\xi}-iG_{o,\xi}\ \ppt{
\hat{b}_\xi+\hat{b}_\xi^{\dagger}
}+\sqrt{2\kappa_{o,\xi}}\ \hat{c}_{o,\xi}^{(in)}\ ,
\\\label{QLE 1b}
\dot{\hat{b}_\xi}(t)&=&-\ppt{
\gamma_{M,\xi}+i\omega_{M,\xi}
}\ \hat{b}_\xi-i\,\ppt{
G_{o,\xi}^{*}\ \hat{c}_{o,\xi}+G_{o,\xi}\ \hat{c}_{o,\xi}^{\dagger}
}
\nn\\&&
-i\,\ppt{
G_{w,\xi}^{*}\ \hat{c}_{w,\xi}+G_{w,\xi}\ \hat{c}_{w,\xi}^{\dagger}
}
+\sqrt{2\gamma_{M,\xi}} \hat{b}_\xi\al{in}\ ,
\end{eqnarray}
where the linearized interaction strengths
\begin{eqnarray}\label{Gixi}
G_{i,\xi}\equiv g_{i,\xi}\,\sqrt{N_{i,\xi}}\ e^{i\varphi_{i,\xi}},
\end{eqnarray}
with $g_{i,\xi}$ the bare interaction strengths, are scaled by the
complex amplitude of the cavity fields $\sqrt{N_{i,\xi}}\ e^{i\varphi_{i,\xi}}$, where $N_{i,\xi}$ are the average number of cavity photons and the phases $\varphi_{i,\xi}$, can be tuned by controlling the phases of the driving fields.
Moreover, $\hat{c}_{i,\xi}^{(in)}$ and $\hat{b}_\xi\al{in}$  are the input noise operators for the cavity modes and the mechanical oscillator. The  cavity input noise operators can be decomposed in terms of the individual decay channels:
$\hat{c}_{i,\xi}^{(in)}(t)=\frac{1}{\sqrt{\kappa_{i,\xi}}}\sum_{j=1}^2\sqrt{\kappa_{i,\xi}\al{j}}\ \hat{c}_{i,\xi}^{(in,j)}(t)$,
where $\hat{c}_{i,\xi}^{(in,j)}$, with  $j\in\pg{1, 2}$, are the quantum input noises with dissipation rates $\kappa_{i,\xi}\al{j}$
for the first and second port of the cavities, such that $\kappa_{i,\xi}=\kappa_{i,\xi}\al{1}+\kappa_{i,\xi}\al{2}$.
All the noise operators (except from the noise operator for the second port of the microwave cavity of the transmitter) are characterized by thermal noise correlations, where
the only non-zero correlation functions are
$\av{b_\xi\al{in}(t)\ b_\xi\al{in}{}\da(t')}=
\av{b_\xi\al{in}{}\da(t)\ b_\xi\al{in}(t')}+\delta(t-t')=
\ppt{1+\bar n_{M,\xi}\al{th}}\delta(t-t')$,
and
$\av{c_{i,\xi}\al{in,j}(t)\ c_{i,\xi}\al{in,j}{}\da(t')}=
\av{c_{i,\xi}\al{in,j}{}\da(t)\ c_{i,\xi}\al{in,j}(t')}+\delta(t-t')=
\ppt{1+\bar n_{i,\xi}\al{th}}\delta(t-t')$, with the number of thermal excitations given by $\bar n_{\iota,\xi}\al{th}=\pq{\exp\ppt{\hbar\,\omega_{\iota,\xi}/k_B\,T}-1}^{-1}$ for $\iota\in\pg{o,w,M}$, which, in particular, is essentially zero for optical frequencies, i.e. $\bar n_{o,\xi}\al{th}=0$.
Instead, the input operator for the second port of the microwave cavity of the transmitter is modified by our feedback system (see Fig.~\ref{Fig1}) as discussed in the next section. We note that we apply the feedback only on the transmitter, and not on the receiver, because, as demonstrated below, the efficiency of the protocol, when properly optimized, is independent from the parameters of the receiver.

\subsection{The electro-optomechanical transmitter within the feedback-loop}

The feedback operates on the EOM transmitter by performing a homodyne detection of the second optical output and by using the corresponding photocurrent to modulate the amplitude of 
the microwave driving field which drives the system through
the second microwave input (see Fig.~\ref{Fig1}). In particular we assume a broadband feedback response function such that the corresponding input noise operator can be expressed as~\cite{feedback-loop sample 1}
\begin{eqnarray}\label{input noise operator w2}
\hat{c}_{w,T}^{(in,2)}(t)=\hat{c}_{w,T}^{(in,2)\circ}(t)+g_{fb}\ \hat{i}_{fb}(t-\tau),
\end{eqnarray}
where $\hat{c}_{w,T}^{(in,2)\circ}$ is the microwave input field without feedback, which is characterized by thermal
noise correlations with
$\bar n_{w,T}\al{th}$ thermal excitations.
Moreover, $g_{fb}$ is the feedback gain and $\hat{i}_{fb}$ is the homodyne photocurrent delayed by the feedback delay time $\tau$. If the delay is sufficiently small it contributes as an additional phase on the expression for the photocurrent, which hence can be approximated as~\cite{feedback-loop sample 1}
\begin{eqnarray}\label{delay time photocurrent}
\hat{i}_{fb}(t-\tau)&=&\sqrt{\frac{\eta}{2}}\pg{ e^{-i\phi}
\pq{\sqrt{2\kappa_{o,T}^{(2)}}\hat{c}_{o,T}(t)-\hat{c}_{o,T}^{(in,2)}(t)} +H.c. }
+\sqrt{1-\eta}\ \hat{X}_{\nu}(t),
\end{eqnarray}
where $\phi\equiv \theta-\Delta_{o,T}\ \tau$, with $\theta$ denoting the homodyne phase, is the modified feedback phase, $\eta$ is the detection efficiency, and $\hat{X}_{\nu}$ accounts for the additional white noise due to the inefficient detection and fulfills the relation $\langle\hat{X}_{\nu}(t)\hat{X}_{\nu}(t^{\prime})\rangle=\frac{1}{2}\delta(t-t^{\prime})$. Using Eqs.~\rp{QLE 1w}, \rp{input noise operator w2} and \rp{delay time photocurrent}, the dynamics of the transmitter's microwave cavity under the effect of a feedback-loop from optical to microwave cavities can be expressed as
\begin{eqnarray}\label{dynamics of w with feedback}
\dot{\hat{c}}_{w,T}(t)&=&-\ppt{\kappa_{w,T}+i\Delta_{w,T}}\,\hat{c}_{w,T}-iG_{w,T} \,\ppt{\hat{b}_T+\hat{b}^{\dagger}_T}
\\&&
+g_{fb}\sqrt{2\,\eta\ \kappa_{o,T}\al{2}\ \kappa_{w,T}\al{2}}
\pq{ e^{-i\phi}\ \hat{c}_{o,T}+e^{i\phi}\ \hat{c}_{o,T}^{\dagger} }+\sqrt{2\,\kappa_{w,T}}\ \hat{c}_{w,T}^{(in)}\nonumber
\end{eqnarray}
where here $\hat{c}_{w,T}^{(in)}$ is the total input noise operator for the microwave mode of the transmitter modified by the feedback, and it is given by (see also the~\ref{App1})
\begin{eqnarray}\label{feedback input noise operator}
\hat{c}_{w,T}^{(in)}&=&\frac{1}{\sqrt{2\kappa_{w,T}}}
\lpq{
\sqrt{2\kappa_{w,T}\al{1}}\ \hat{c}_{w,T}^{(in,1)}+
\sqrt{2\kappa_{w,T}\al{2}}\ \hat{c}_{w,T}^{(in,2)\circ}
}\nn\\&& \rpq{
-g_{fb}\ \sqrt{\eta\ \kappa_{w,T}\al{2}}\ \ppt{
e^{-i\phi}\ \hat{c}_{o,T}^{(in,2)}+e^{i\phi}\ \hat{c}_{o,T}^{(in,2)\dagger}
}
+g_{fb}\ \sqrt{2\,\kappa_{w,T}\al{2}(1-\eta)}\ \hat{X}_{\nu}
}\ .
\end{eqnarray}

\subsection{The system in the rotating wave approximation and the output fields}\label{Sec.RWA}

Both EOM systems are operated by driving the microwave cavity on the red sideband and the optical cavity on the blue one, i.e. $\Delta_{w,\xi}=-\Delta_{o,\xi}=\omega_{M,\xi}$.
If the interaction strengths are not too large, it is possible to perform a rotating wave approximation by retaining only the resonant terms.
In particular,
by rewriting the QLEs [Eqs.~\rp{QLE 1w}-\rp{QLE 1b}, \rp{dynamics of w with feedback}] in the interaction picture with respect to the free Hamiltonian, they can be reduced to the form
(see~\ref{App1})
\begin{eqnarray}\label{QLEs}
\dot{\hat{c}}_{w,\xi}&=& -\kappa_{w,\xi}\ \hat{c}_{w,\xi}-i\,G_{w,\xi}\ \hat{b}_\xi
+\mu_\xi\
\hat{c}_{o,\xi}^{ \dagger}
+\sqrt{2\ \kappa_{w,\xi}}\ \hat{c}_{w,\xi}^{(in)},
\nn\\
\dot{\hat{c}}_{o,\xi}&=& -\kappa_{o,\xi}\ \hat{c}_{o,\xi}-i\,G_{o,\xi}\ \hat{b}_\xi^{ \dagger}+\sqrt{2\ \kappa_{o,\xi}}\ \hat{c}_{o,\xi}^{(in)},
\nn\\
\dot{\hat{b}}_\xi&=& -\gamma_{M,\xi}\ \hat{b}_\xi-i\,G_{o,\xi}\ \hat{c}_{o,\xi}^{\dagger}-i\,G_{w,\xi}^{*}\ \hat{c}_{w,\xi}+\sqrt{2\ \gamma_{M,\xi}}\ \hat{b}_{\xi}\al{in}.
\end{eqnarray}
where
\begin{eqnarray}\label{muxi}
\mu_T&=&g_{fb}\,\sqrt{2\eta\ \kappa_{o,T}\al{2}\ \kappa_{w,T}\al{2}}\ e^{i\phi}
\nn\\
\mu_R&=&0
\end{eqnarray}
which indicate that the feedback is applied only on the transmitter.
Moreover, in the rotating wave approximation, the microwave input for the transmitter describes thermal noise with a number of excitations modified by the feedback according to the relation $\bar n_{w,T}\al{fb}=\bar n_{w,T}\al{th}+g_{fb}^2\,\kappa_{w,T}\al{2}/2\,\kappa_{w,T}$ (see~\ref{App1}).

The QLEs can be easily solved in the Fourier space and, using the standard input-output relations $\hat c_{i,\xi}\al{out,j}=\sqrt{2\,\kappa_{i,\xi}\al{j}}\ \hat c_{i,\xi}-\hat c_{i,\xi}\al{in,j}$, the output propagating modes from the EOM systems can be expressed in terms of the input noise operators. In particular, here we are interested in the first microwave and optical outputs of the transmitter, which are, respectively, the signal and the idler of the QI protocol, and the first optical output of the receiver that is the converted signal, which will be combined with the idler and eventually detected (see Fig.~\ref{Fig1}).
In very general terms the output modes, in Fourier space, can be expressed as linear combinations of the input noise operators
\begin{eqnarray}\label{Final output field (o,1)}
c_{o,\xi}^{(out,1)}(\omega)
&=&
\mathcal{A}_\xi(\omega)\ b_{\xi}\al{in}{}^{\dagger}(\omega)
+\mathcal{B}_\xi(\omega)\ c_{w,\xi}^{(in,1)\dagger}(\omega)
+\mathcal{C}_\xi(\omega)\ c_{w,\xi}^{(in,2)\dagger}(\omega)
\nn\\&&
+\mathcal{D}_\xi(\omega)c_{o,\xi}^{(in,1)}(\omega)
+\mathcal{E}_\xi(\omega)c_{o,\xi}^{(in,2)}(\omega),
\end{eqnarray}
and
\begin{eqnarray}\label{Final output field (w,1)}
c_{w,\xi}^{(out,1)}(\omega)
&=&
\mathcal{A}_\xi^{\prime}(\omega)\ b_{\xi}\al{in}(\omega)
+\mathcal{B}_\xi^{\prime}(\omega)\ c_{w,\xi}^{(in,1)}(\omega)
+\mathcal{C}_\xi^{\prime}(\omega)\ c_{w,\xi}^{(in,2)}(\omega)
\nn\\&&
+\mathcal{D}_\xi^{\prime}(\omega)\ c_{o,\xi}^{(in,1)\dagger}(\omega)
+\mathcal{E}_\xi^{\prime}(\omega)\ c_{o,\xi}^{(in,2)\dagger}(\omega),
\end{eqnarray}
where the explicit expressions for the coefficients corresponding to the different inputs are reported in~\ref{App2}.
In QI the signal and the idler are entangled. In general, only spectral modes at opposite frequencies (with respect to the corresponding cavity resonance frequency) of Gaussian stationary continuous-wave fields can be entangled~\cite{zippilli2015}. For this reason here we focus on
the microwave mode  $c_{w,\xi}^{(out,1)}(\omega)$
at frequency $\omega$ and
the optical mode $c_{o,\xi}^{(out,1)}(-\omega)$
at the opposite frequency $-\omega$. The state of these two modes is fully characterized by the number of photons of the two modes, defined by the relation
\begin{eqnarray}\label{n}
\av{c_{i,\xi}^{(out,1)}{}\da(\omega)\ c_{i,\xi}^{(out,1)}(\omega')}=n_{i,\xi}(\omega)\ \delta(\omega+\omega')\ ,
\end{eqnarray}
and by their correlation
\begin{eqnarray}\label{m}
\av{c_{w,\xi}^{(out,1)}(\omega)\ c_{o,\xi}^{(out,1)}(\omega')}=m_\xi(\omega)\ \delta(\omega+\omega')\ .
\end{eqnarray}
In the following, even when we omit, for simplicity, the frequency argument, the
microwave
modes have to be intended at frequency $\omega$ and the
optical
ones at the opposite frequency $-\omega$.

\section{Microwave Quantum Illumination}\label{Sec.MQI}

In microwave QI the EOM receiver is used to convert and phase conjugate the returned microwave signal to the optical regime where it can be combined with the idler and eventually efficiently detected (note that a phase-conjugate receiver operated fully in the optical regime has been alredy experimentally demonstrated in Ref.~\cite{hao2021}).

Here, the returned microwave signal is taken as input to the first input port of the microwave cavity of the receiver (see Fig.~\ref{Fig1}).
In particular, one identifies two scenarios: in the absence of the target (hypothesis $H_0$) the returned field is made just by background photons, described by the annihilation operator $c_B$ and characterized by $\av{c_B\da\ c_B}=N_B$ average photons. Thus, in this case the first microwave input for the receiver is simply given by $c_{w,R}\al{in,1}=c_B$.
Instead, in the presence of a target (hypothesis $H_1$) with low reflectivity $t\ll1$, the returned signal is a superposition of background and reflected photons, and hence $c_{w,R}\al{in,1}=\sqrt{t}\ c_{w,T}\al{out,1}+\sqrt{1-t}\ c_B$, where now $\av{c_B\da\ c_B}=N_B/(1-t)$ such that the number of background photons at the receiver is the same under both hypotheses~\cite{QI with Gaussian states}.

The EOM receiver described in Sec.~\ref{system} is able to convert and phase-conjugate the first microwave input. To be specific, as shown in Eq.~\rp{Final output field (o,1)}, one finds that $c_{o,R}\al{out,1}\propto c_{w,R}\al{in,1}{}\da +$ [contributions from other input noise operators]. Then, as in Ref.~\cite{Microwave QI 1}, the detection of the QI protocol follows the procedure of the phase conjugated receiver of Ref.~\cite{Gaussian state QI Guha} which, even though not optimal, is known to provide 3 dB gain in signal-to-noise-ratio, and it has been realized by means of digital post-processing in Ref.~\cite{Microwave QI 2}. The optical output $c_{o,R}\al{out,1}$ is mixed with the idler $c_{o,T}\al{out,1}$ on a 50/50 beam splitter and finally the two fields after the beam splitter (described by the annihilation operators $d_\pm=\pq{c_{o,R}\al{out,1}\pm c_{o,T}\al{out,1}}/\sqrt{2}$) are measured by direct photodetection and the detected photon numbers $N_\pm=d_\pm\da\ d_\pm$ are eventually subtracted. The resulting quantity $N=N_+-N_-$ can be used to discriminate the presence of the target~\cite{Gaussian state QI Guha,Microwave QI 1,Microwave QI 2}. In particular the QI protocol makes use of $M$ identical and independent copies of entangled signal-idler pairs. 
In the limit of large $M$\footnote{
Note that, in general, $M$ is given by the product of the characteristic bandwidth of the system and the time. So a sufficiently large $M$ can be achieved using a sufficiently large time. In our case the bandwidth is defined by the decay rate of our system, determined by the minimum real parts of the eigenvalues of the matrix of coefficients of the quantum Langevin equations~\rp{QLEs}. In the case of the results of presented in the next section these decay rates are of the order of $10^4$--$10^5$. This means that a sufficiently large M (of the order of $10^4$--$10^5$) is realized for a time of the order of one second.}, the corresponding error probability $P_{err}$ in discriminating the two hypothesis $H_0$ and $H_1$ can be expressed as $P_{err}={\rm erfc}\ppt{\sqrt{SNR_{QI}\al{M}/8}}/2$ in terms of the
complementary error function erfc, and of the signal to noise ratio
$SNR_{QI}\al{M}=4\, M\ppt{N\bigl|_{H_1}-N\bigl|_{H_0}}^2/\ppt{\sigma\bigl|_{H_1}+\sigma\bigl|_{H_0}}^2$,
with $N\bigl|_{H_j}$ and $\sigma\bigl|_{H_j}$ average and variance of the variable $N$ under hypothesis $H_j$, respectively~\cite{Microwave QI 1}.

The signal to noise ratio for microwave QI with our system ($SNR_{QI}\al{M}$) can be easily computed using Eqs.~\rp{Final output field (o,1)} and \rp{Final output field (w,1)}. In the limit of large number of background photons $N_B$, it can be approximated by the expression (see \ref{App3})
\begin{eqnarray}\label{QI SNR}
SNR_{QI}^{(M)}\simeq\frac{4\ M\ t\ \pg{ Re [\mathcal{B}_R(-\omega)\ m_T^{*}(\omega)]}^2}{\vert \mathcal{B}_R(-\omega)\vert^2\ N_B\ \pq{1+2\ n_{o,T}(-\omega)}},
\end{eqnarray}
where $\mathcal{B}_R(\omega)$ is the coefficient corresponding to the first microwave input noise operator (see Eq.~\rp{Final output field (o,1)}], and $n_{o,T}(-\omega)$ and $m_T(\omega)$ are, respectively, the idler number of excitations and the
idler--signal correlation defined in Eqs.~\rp{n} and \rp{m}.
Evidently, at fixed amplitude of the parameters $\BBB_R(-\omega)$ and $m_T^*(\omega)$, the maximum of the signal-to-noise ratio is achieved when the relative phase between these two parameters is selected such that $Im\big [m_T^{*}(\omega)\mathcal{B}_R(-\omega)\big]=0$. In this case one finds that the optimal signal-to-noise ratio is given by
\begin{eqnarray}\label{optSNR}
SNR_{QI,{\rm opt}}^{(M)}=\frac{4\ M\ t\ \vert m_T(\omega)\ \vert^2}{N_B\pq{1+2\ n_{o,T}(-\omega)}}\ .
\end{eqnarray}
As shown in~\ref{App2} the parameter $\BBB_R(-\omega)$ is proportional to the product of the optomechanical coupling strengths $G_{o,R}\ G_{w,R}$, the phases of which can be controlled by controlling the phases of the driving fields. By this means one can always select the phase of $\BBB_R$ which realizes the optimal signal-to-noise ratio~\rp{optSNR}.
Specifically, in order to optimize the QI protocol, in the numerical results presented below, we set the phases of the driving fields on the receiver [see Eq.~\rp{Gixi}] according to the relation
\begin{eqnarray}\label{phase matching condition}
\ee^{-\ii\ppt{\varphi_{w,T}+\varphi_{o,T}}}=
\frac{\BBB_R(-\omega)\ m_T^*(\omega)}{G_{o,R}\ G_{w,R}}
\abs{\frac{G_{o,R}\ G_{w,R}}{\BBB_R(-\omega)\ m_T^*(\omega)}}\ .
\end{eqnarray}
It is important to note that Eq.~\rp{optSNR} does not depend on the parameters of the receiver. This means that in the limit of strong background noise, once the driving phases are properly selected, it is not possible to improve the microwave QI performance by further tweaking the receiver. This is the reason why, in this work, we have not included a feedback system also on the receiver.

The importance of QI is that it can outperform any classical protocol which employs the same number of signal photons. In order to have a clear understanding of the efficiency of QI is, therefore, necessary to compare QI with the optimal classical protocol. Here, analogously to the analysis of Ref.~\cite{Microwave QI 1}, we compare the signal-to-noise ratio in the quantum case ($SNR_{QI}\al{M}$), with the signal-to-noise ratio achievable with the classical benchmark of homodyne detection of the returned signal after transmission of a coherent state, with $n_{w,T}$ photons, which is given by $SNR_{Cl}\al{M}=4\,M\,t\,n_{w,T}/\pq{2(N_B+1)}$~\cite{QI with Gaussian states, Gaussian state QI Guha}.
In particular, here we analyze the ratio
\begin{eqnarray}\label{FF}
\FF=\frac{SNR_{QI}\al{M}}{SNR_{Cl}\al{M}}
\end{eqnarray}
between the two signal-to-noise ratios as a function of the system parameters. A value of $\FF$ larger than one indicates that the QI is effective and outperforms the optimal classical protocol employing the same number of photons.

We also note that in the optimal case discussed in Eq.~\rp{optSNR}, the ratio $\FF$ reduces to
\begin{eqnarray} \label{eq:fopt}
\FF_{opt}=\frac{2\ \abs{m_T(\omega)}^2}{n_{w,T}(\omega)\pq{1+2\ n_{o,T}(-\omega)}}\ .
\end{eqnarray}
Since, for a two-mode entangled state, with excitation numbers of the two modes $n_w$ and $n_o$, the two-mode correlation $m$ is constrained by the relation $\abs{m}\leq\sqrt{{\rm min}\pg{ n_w ( 1 + n_o ) , n_o ( 1 + n_w ) }}$, the maximum ratio is achieved for $n_{w,T}\leq n_{o,T}$ and it is given by
\begin{equation}\FF_{max}=1+ \frac{1}{1 + 2\, n_{o,T}}.\end{equation}
This shows that the best result for QI is achieved when the correlation $|m_T|$ is as large as possible, but together with the smallest value of photons.
We also verify that the ratio $\FF$ can never surpass the maximum value of $2$, consistently with the properties of the phase-conjugated receiver~\cite{Gaussian state QI Guha} employed here. More generally speaking, we recall that, according to the Helstrom bound~\cite{Helstrom}
(see also Refs.~\cite{Wilde2017,DePalma2018,Nair2020}) which, in the limit of very large $M$, asymptotically coincides with the Chernoff bound~\cite{QI with Gaussian states, Gaussian state QI Guha,Pirandola2008}, the ratio $\FF$ can never surpass the value of $4$. 
In principle, this upper limit could be achieved by the receiver of Ref.~\cite{Zhuang}, which is however prohibitively hard to realize experimentally (see also Refs.~\cite{Zhuang2017,Zhuang2020} for other related studies).
In any case, we note that the improved performance of the QI protocol, which we demonstrate hereafter, is achieved by means of the feedback which improves the preparation of the entangled resource (i.e. signal-idler entangled state). For this reason, we expect that a similar improvement of the protocol can be observed also for different and more refined detection strategies. Moreover Eq.~\rp{eq:fopt} also means that maximum entanglement (which can be achieved for large number of photons) does not necessarily correspond to the best performance of the QI protocol, compared to the classical benchmark.

A final comment is in order. The analysis that we have performed is based on the assumption that we can detect a single spectral component of the signal and of the idler at the appropriate frequencies. This
can always be done by filtering the fields with, for example, very narrow Fabry-Perot cavities, and then detecting them with photodetectors with long enough integration times.

\section{Numerical results}\label{num}

\begin{figure}\centering
\begin{center}
\includegraphics[width=0.76\textwidth]{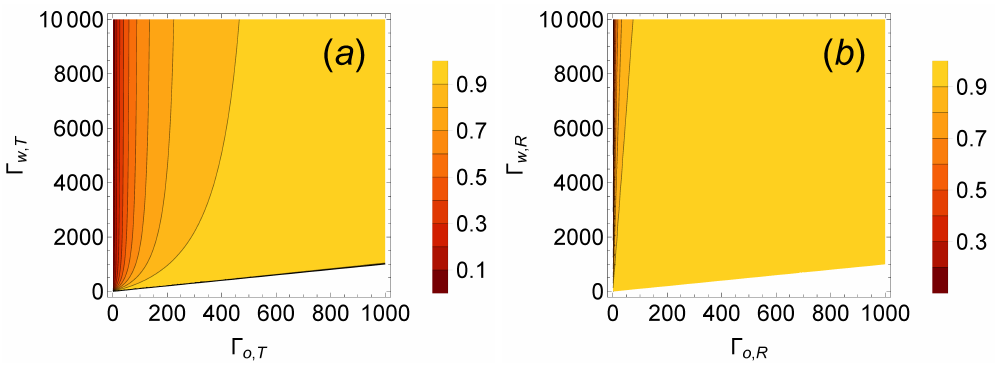}
\end{center}
\caption{Ratio $\mathcal{F}$ [see Eq.~\rp{FF}], versus the opto- and electromechanical cooperativities [$\Gamma_{i,\xi}=G_{i,\xi}^2/(\gamma_{M,\xi}\, \kappa_{i,\xi})$, for $i\in\pg{o,w}$ and $\xi\in\pg{T,R}$] of the transmitter $(a)$ and receiver $(b)$, without feedback ($g_{fb}=0$). The cavities are two-sided and symmetric with $\kappa_{i,\xi}\al{1}=\kappa_{i,\xi}\al{2}=\kappa_{i,\xi}/2$.
In $(a)$
$ \Gamma_{o,R}=995, \Gamma_{w,R}=1120$. In $(b)$  $\Gamma_{o,T}=1000, \Gamma_{w,T}=1440$.
Both plots are obtained for a $10$-ng-mass mechanical resonator with $\omega_{M,\xi}/2\pi=10 MHz$ and $Q=30 \times 10^3$, at a temperature of $30 mK$; a microwave cavity with $\omega_{w,\xi}/2\pi=10GHz$ and $\kappa_{w,\xi}=0.2\omega_M$; and an
optical cavity with $\kappa_{o,\xi}=0.1\omega_M$.
The opto-mechanical and electro-mechanical coupling rates are $g_{o,\xi}/2\pi=115.512 Hz$ and $g_{w,\xi}/2\pi=0.327 Hz$. The target reflectivity is $t=0.07$.
The white areas indicate the parameter regimes in which the system is unstable.}
\label{Fig2}
\end{figure}

\begin{figure}\centering
\includegraphics[width=0.96\textwidth]{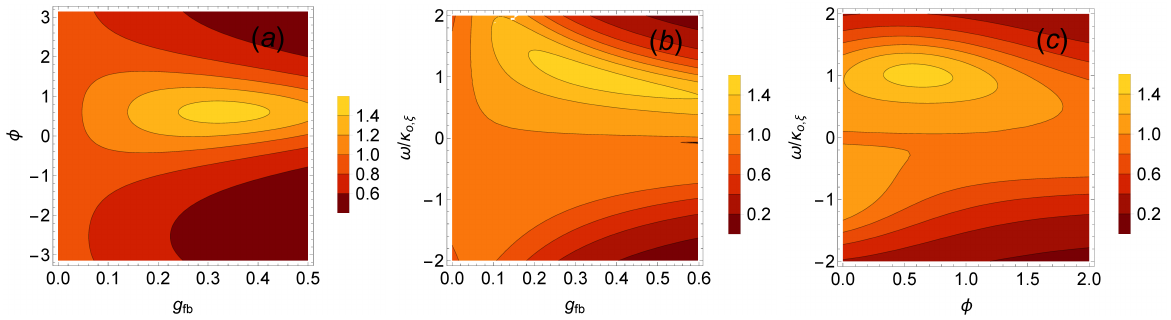}
\caption{(Color online)
Ratio $\mathcal{F}$ versus the feedback parameters, $g_{fb}$ and $\phi$ and the frequency.
In (a) $\omega=1.014\,\kappa_{o,\xi}$. In (b) $\phi=0.6$. In (c) $g_{fb}=0.33$.
In all the plots $\Gamma_{o,T}=1000$, $\Gamma_{w,T}=1440$, $\Gamma_{o,R}=995$, $\Gamma_{w,R}=1120$; The feedback detection efficiency is $\eta=1$; The phases of the coupling strengths are chosen in order to fulfill Eq. (\ref{phase matching condition}); The other parameters are as in Fig.~\ref{Fig2}.}
\label{Fig3}
\end{figure}

In all the results presented hereafter we have adjusted the phase of the driving fields such that the optimal condition defined by Eq.~\rp{phase matching condition} is always fulfilled.

\subsection{Symmetric cavities}

As a preliminary result, we demonstrate in Fig~\ref{Fig2} that when there is no feedback the QI with the present system is inefficient (i.e. $\FF$ is below one). This is due to the fact that here we employ two-sided symmetric (same decay rate for the two decay channels) cavities. These results should be compared with Ref.~\cite{Microwave QI 1}
where efficient microwave QI is demonstrated with an analogous setup which employs single-sided cavities. Here the additional noise due to the additional decay channels inhibits the performance of our system.

In order to study the effect of feedback, we have initially
maximized numerically the ratio $\FF$ as a function of the feedback parameters $g_{fb}$ and $\phi$, the cooperativities $\Gamma_{i,\xi}=G_{i,\xi}^2/(\gamma_{M,\xi}\, \kappa_{i,\xi})$  (for $i=o,w$ and $\xi=T,R$) and the frequency $\omega$. In this way, we have found a set of parameters for which $\FF$ reaches values comparable with those discussed in Ref.~\cite{Microwave QI 1}.
In Fig.~\ref{Fig3} we plot $\FF$ as a function of the feedback parameters and of the frequency when the other parameters, in each plot, are set to the optimal values obtained by the initial maximization.
This figure shows that $\FF$ can reach values well above one. Thus,
it demonstrates that by properly tuning the feedback system, the QI protocol can surpass
the efficiency achievable in the classical case also when the presence of photon losses would have prevented it.
Fig.~\ref{Fig3} (b) and (c) also show that, when we use feedback, the optimum QI is obtained for $\omega\neq 0$, that is, off-resonance with respect to the microwave and optical cavities of the system, differently from what occurs in the case of Ref.~\cite{Microwave QI 1}.

\subsection{Asymmetric cavities}

\begin{figure}\centering
\hspace*{-1cm}
\begin{minipage}{18cm}\centering
\includegraphics[width=0.96\textwidth]{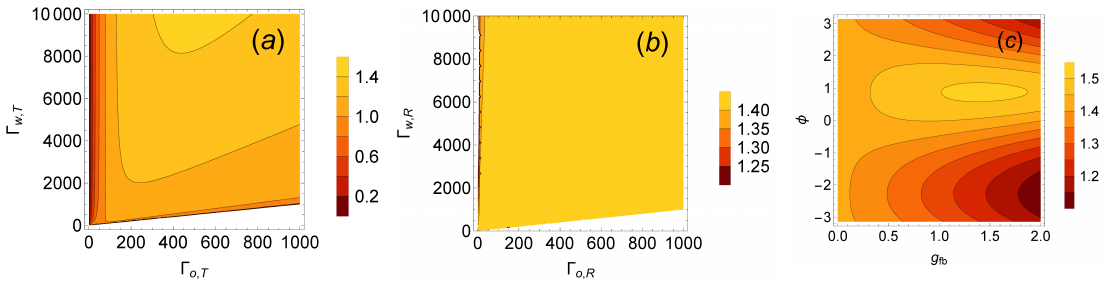}
\end{minipage}
\caption{(Color online)
Ratio, $\mathcal{F}$,  versus the opto- and electromechanical cooperativities of the transmitter $(a)$ and receiver $(b)$ and versus the feedback parameters (c), when the cavities are asymmetric, with larger dissipation on the first sides: $\kappa_{i,\xi}\al{1}=0.9\kappa_{i,\xi}$, and $\kappa_{i,\xi}\al{2}=0.1\kappa_{i,\xi}$, for $i\in\pg{o, w}$  and $\xi\in\pg{T, R}$.
(a) and (b) are with no feedback.
In (a) 
In (a) $\omega/\kappa_{o,\xi}=1.8 \times 10^{-4}$,  $\Gamma_{o,R}=200.5$, $\Gamma_{w,R}=553.17$; in (b)  $\omega/\kappa_{o,\xi}=1.8 \times 10^{-4}$,  $\Gamma_{o,T}=480.4$, $\Gamma_{w,T}=10000$; in (c)  $\omega/\kappa_{o,\xi}=0.473$,  $\Gamma_{o,T}=480.4$, $\Gamma_{w,T}=10000$, $\Gamma_{o,R}=200.5$, $\Gamma_{w,R}=553.17$.
The other parameters are as Fig.~\ref{Fig3}.
}
\label{Fig5}
\end{figure}

We have performed a similar analysis also in the case of asymmetric cavities (see Fig.~\ref{Fig5}). Specifically, we consider the situation in which the decay rate through the second mirror (the one used for the feedback)
is much smaller than the decay rate through the first. As before, we first consider the case without feedback [see Figs.~\ref{Fig5} (a) and (b)]. In this situation the system (without feedback) is similar to the single-sided cavities of Ref.~\cite{Microwave QI 1} with only small additional dissipation, and the value of $\FF$ is larger than one. This indicates that the protocol is resilient to weak added noise.
In this case the feedback still allows to improve the performance, as shown in Fig.~\ref{Fig5} (c), but the improvement is significantly smaller than in the symmetric case.
In fact, we have verified that, when we use feedback with two-sided cavities, the value of $\FF$  never reaches the maximum value achievable with single-sided cavities (without feedback) and equal total dissipation rate.

\subsection{QI and entanglement}

\begin{figure}\centering
\includegraphics[width=0.64\textwidth]{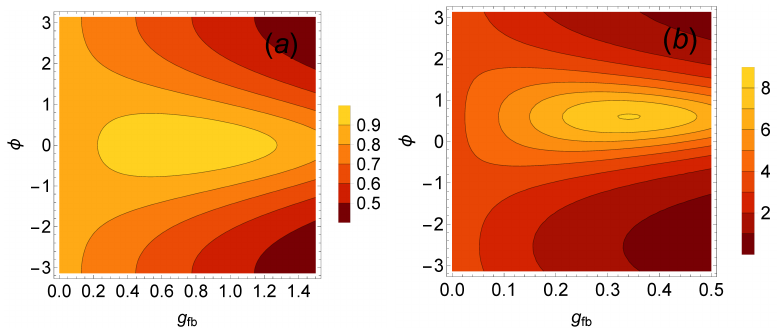}
\caption{(Color online)
(a) Logarithmic Negativity, $E_N$, and (b) normalized Logarithmic negativity, $E_N/n_{w,T}$, versus the feedback-loop parameters, $g_{fb}$ and $\phi$. It has been assumed that there is a feedback-loop with perfect detection efficiency, $\eta=1$, from the two-sided symmetric optical to microwave cavity of the EOM system. The selected frequency as well as optomechanical and electromechanical cooperativities are respectively $\omega/\kappa_{o,T}=-0.0033$, $\Gamma_{o,T}=1000$ and $\Gamma_{w,T}=3403$, which along with the proper values of the feedback parameters maximize $E_N$. See Fig. (\ref{Fig3}) for other parameter values.}
\label{Fig8}
\end{figure}

QI exploits entanglement to outperform classical protocols for the detection of weakly reflecting targets in bright noisy backgrounds.
It is, thus, interesting to analyze the precise relation between entanglement and QI performance.
As already discussed in Sec.~\ref{Sec.MQI} we expect that optimum QI may not correspond to maximum entanglement. This is indeed described by Fig.~\ref{Fig8}. In plot (a) we report the signal-idler entanglement at the output of the transmitter, measured in terms of the logarithmic negativity~\cite{zippilli2015}, corresponding to the results of Fig.~\ref{Fig3} (a). We note that, in this system, the entanglement is only weakly enhanced by the feedback [see Fig.~\ref{Fig8} (a)]. Moreover the enhancement is observed for zero feedback phase. On the other hand, we have seen in Fig.~\ref{Fig3} (a) a significant improvement of the QI protocol with a maximum at a finite value of the feedback phase. In fact, as discussed at the end of Sec.~\ref{Sec.MQI}, we expect to observe more efficient QI when the entanglement is maximum for the smallest possible number of excitations. This is, indeed, shown in Fig.~\ref{Fig8} (b) where we plot the logarithmic negativity rescaled by the number of signal photons, $E_N/n_{w,T}$. Here it is evident that this normalized logarithmic negativity follows precisely the behavior of the ratio $\FF$ in Fig.~\ref{Fig3} (a).

\subsection{Effect of the feedback detection efficiency}

\begin{figure}[b]\centering
\includegraphics[width=0.4\textwidth]{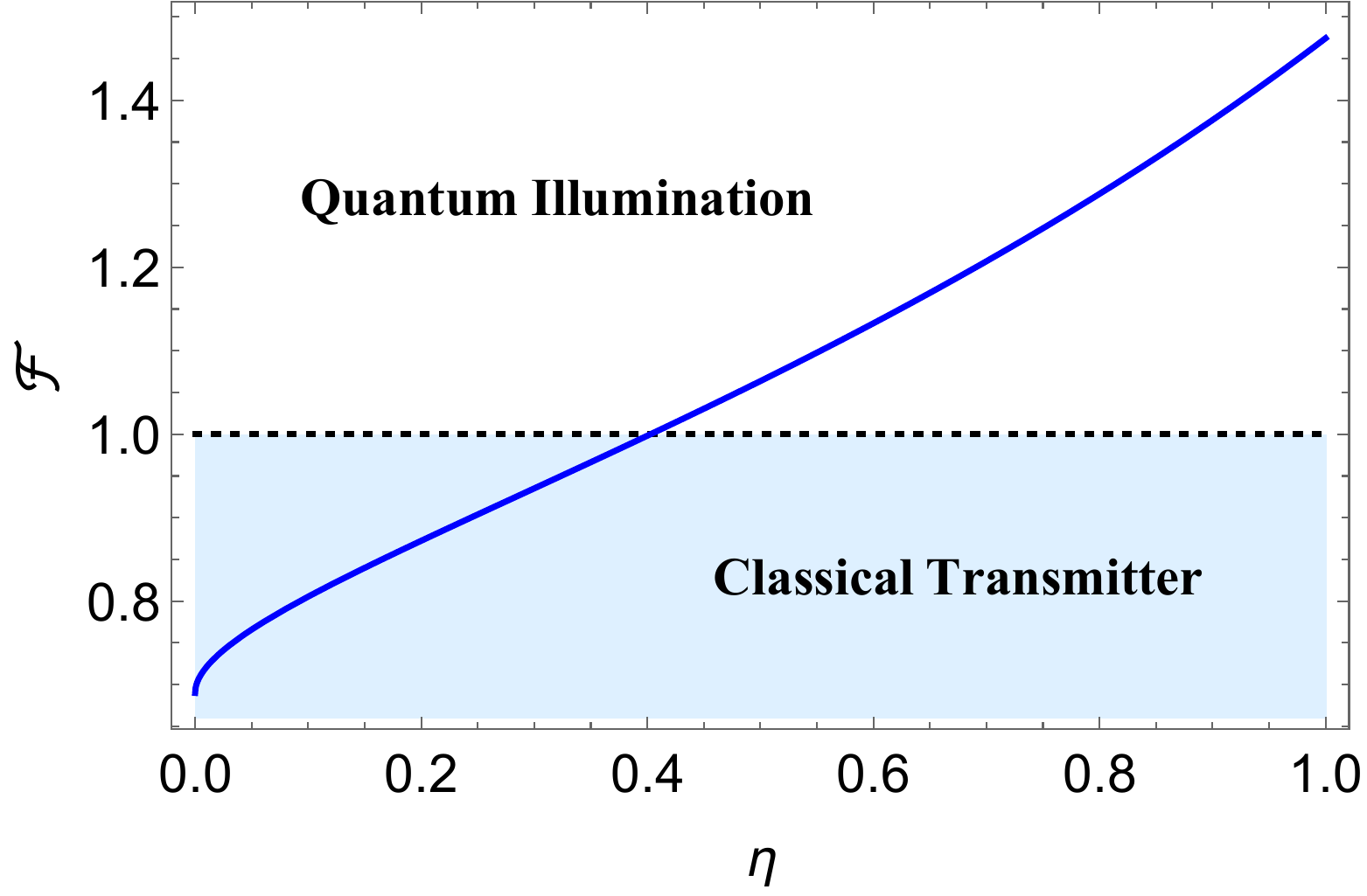}
\caption{(Color online) Ratio $\mathcal{F}$, versus the detection efficiency, $\eta$.
The other parameters are as in Fig.~\ref{Fig3}.}
\label{Fig12}
\end{figure}

All the results presented so far are obtained with perfect feedback detection efficiency, i.e. $\eta=1$ [see Eq.~\rp{delay time photocurrent}]. Inefficient detection entails additional detection noise. It is therefore important to assess accurately the effect of this additional noise source on the QI protocol. This is done in Fig.~\ref{Fig12}. Here we consider symmetric cavities with the parameters of Fig.~\ref{Fig3} (a), and we focus on the feedback gain and phase which maximize the value of $\FF$. The plot shows  that the protocol is efficient also for relatively low values of $\eta$.

\section{Conclusions}\label{concl}

In conclusion we have designed and analyzed a feedback setup which acts on the transmitter of a microwave quantum illumination device based on two EOM systems and first proposed in Ref.~\cite{Microwave QI 1}. We have shown that this feedback-enhanced QI device can operate efficiently also in regimes where, in absence of feedback, excessive dissipation inhibits the performance of the QI protocol.
%

In particular we have considered electro-opto-mechanical devices constituted of two-sided electromagnetic cavities such that QI with these systems, based on the approach of Ref.~\cite{Microwave QI 1}, is not able to outperform the optimal classical protocol. By employing a feedback loop which measures the unused output of the optical cavity, and acts on the field driving the microwave cavity, it is possible to recover part of the quantum correlation lost in these unused output and, hence, to improve the quality of the generated entanglement. As a consequence, one can achieve efficient microwave quantum illumination outperforming every classical protocol with the same number of signal photons.
We have also analyzed in details the effect of the feedback detection efficiency, thereby demonstrating that the protocol is particularly resilient to detection inefficiencies, showing once again that homodyne-feedback schemes can be tuned to improve the performance of quantum optical systems~\cite{feedback-loop sample 1, feedback-loop sample 2, feedback-loop sample 3}.

\ack
We acknowledge the support of the European Union Horizon 2020 Programme for Research and Innovation through the Project No. 862644 (FET Open QUARTET).

\appendix
\setcounter{section}{0}

\section{ -- The Model in the rotating wave approximation}
\label{App1}

Here we describe how to derive the approximated quantum Langevin equations~\rp{QLEs} starting form the full linearized model described by Eqs.~\rp{QLE 1w}-\rp{QLE 1b} and \rp{dynamics of w with feedback}.
In particular,
Eq.~\rp{QLEs} is found by first rewriting Eqs.~\rp{QLE 1w}-\rp{QLE 1b} and \rp{dynamics of w with feedback} in the interaction picture with respect to the free system Hamiltonian, and then eliminating the non resonant terms, under the assumption that the interaction strengths are sufficiently small, i.e.
\begin{eqnarray}\label{conditionRWA}
\vert G_{w,\xi}\vert ,\ \vert G_{o,\xi}\vert,\
\abs{\mu_\xi}
\ll \omega_{M,\xi},\ \abs{\Delta_{i,\xi}}\ ,
\end{eqnarray}
where $\mu_\xi$ is defined in Eq.~\rp{muxi}.
Specifically, we employ the transformations
\begin{eqnarray}\label{transformation}
\hat{c}_{w,\xi}(t)&=&e^{-i\,\Delta_{w,\xi}\, t}\ \hat{c}_{w,\xi}^{\prime}(t),\\
\hat{c}_{o,\xi}(t)&=& e^{-i\,\Delta_{o,\xi}\, t}\ \hat{c}_{o,\xi}^{\prime}(t),\\
\hat{b}_\xi(t)&=& e^{-i\,\omega_{M,\xi}\ t}\, \hat{b}_\xi^{\prime}(t)\ ,
\end{eqnarray}
and we obtain the equations for the slowly varying operators
\begin{eqnarray}
\label{QLE 3}
\dot{\hat{c}}_{w,\xi}^{\prime}(t)&=&-\kappa_{w,\xi}\ \hat{c}_{w,\xi}^{\prime} -i\,G_{w,\xi}\
\pq{
\hat{b}_\xi^{\prime}\ e^{i\,(\Delta_{w,\xi}-\omega_{M,\xi})\,t}+
\hat{b}_\xi^{\prime \dagger}\ e^{i\,(\Delta_{w,\xi}+\omega_{M,\xi})\,t}
}
\nn\\&&
+\abs{\mu_\xi}
\pq{
e^{-i\,\phi}\ \hat{c}_{o,\xi}^{\prime}\ e^{i\,(\Delta_{w,\xi}-\Delta_{o,\xi})\,t}+e^{i\,\phi}\ \hat{c}_{o,\xi}^{\prime \dagger}\  e^{i\,(\Delta_{w,\xi}+\Delta_{o,\xi})\,t}
}
+\sqrt{2\,\kappa_{w,\xi}}\ \hat{c}_{w,\xi}^{\prime\,(in)}\ ,
\nn\\
\dot{\hat{c}}_{o,\xi}^{\prime }(t)&=&-\kappa_{o,\xi}\ \hat{c}_{o,\xi}^{\prime}-i\,G_{o,\xi}\
\pq{
\hat{b}_\xi^{\prime}\ e^{i\,(\Delta_{o,\xi}-\omega_{M,\xi})\,t}+\hat{b}_\xi^{\prime \dagger}\ e^{i\,(\Delta_{o,\xi}+\omega_{M,\xi})\,t}
}
+\sqrt{2\,\kappa_{o,\xi}}\ \hat{c}_{o,\xi}^{\prime\, (in)}\ ,
\nn\\
\dot{\hat{b}}_\xi^{\prime}(t)&=& -\gamma_{M,\xi}\ \hat{b}_\xi^{\prime}-
i\,\pq{
G_{o,\xi}^{*}\ \hat{c}_{o,\xi}^{\prime}\ e^{i\,(\omega_{M,\xi}-\Delta_{o,\xi})\,t}+
G_{o,\xi}\ \hat{c}_{o,\xi}^{\prime \dagger}\ e^{i\, (\omega_{M,\xi}+\Delta_{o,\xi})\,t}
}
\nn\\&&
-i \pq{
G_{w,\xi}^{*}\ \hat{c}_{w,\xi}^{\prime}\ e^{i\,(\omega_{M,\xi}-\Delta_{w,\xi})\,t}+
G_{w,\xi}\ \hat{c}_{w,\xi}^{\prime \dagger}\ e^{i\,(\omega_{M,\xi}+\Delta_{w,\xi})\,t}
}
+\sqrt{2\,\gamma_{M,\xi}}\ \hat{b}_\xi^{\prime\,(in)}\ ,
\end{eqnarray}
where we have
introduced the rotated input noise operators $\hat{c}_{i,\xi}^{\prime\,(in)}=\hat{c}_{i,\xi}^{(in)}\ e^{i\,\Delta_{i,\xi}\,t}$
and
$\hat{b}_{\xi}^{\prime\,(in)}=\hat{b}_{\xi}^{(in)}\ e^{i\,\omega_{M,\xi}\,t}$.
Now, assuming the resonance condition $\Delta_{w,\xi}=-\Delta_{o,\xi}=\omega_{M,\xi}$ discussed at the beginning of Sec.~\ref{Sec.RWA}, we find that the only resonant terms in these equations are those which constitute Eq.~\rp{QLEs} (where, in order to ease the notation, we have removed the ``prime'' symbol from the operators).
We also note that the correlation functions of
the feedback-modified microwave input noise operator of the transmitter, defined in Eq.~\rp{feedback input noise operator}, are given by
\begin{eqnarray}\label{correlation functions of the feedback-modified noise operator 2}
\av{\hat{c}_{w,T}^{(in)}{}\da(t)\ \hat{c}_{w,T}^{(in)}{}\da(t')}&=&
\av{\hat{c}_{w,T}^{(in)}(t)\ \hat{c}_{w,T}^{(in)}(t')}=\frac{g_{fb}^2\,\kappa_{w,T}\al{2}}{2\,\kappa_{w,T}}\ \delta(t-t')\ ,
\\
\av{\hat{c}_{w,T}^{(in)}(t)\ \hat{c}_{w,T}^{(in)}{}\da(t')}&=&
\av{\hat{c}_{w,T}^{(in)}{}\da(t)\ \hat{c}_{w,T}^{(in)}(t')}+\delta(t-t')
=\ppt{1+\bar n_{w,T}\al{fb}}\,\delta(t-t')\ ,
\nn
\end{eqnarray}
where
\begin{eqnarray}
\bar n_{w,T}\al{fb}=\bar n_{w,T}\al{th}+\frac{g_{fb}^2\,\kappa_{w,T}\al{2}}{2\,\kappa_{w,T}}\ .
\end{eqnarray}
These correlations in general describe classical squeezed thermal noise, however, when Eq.~\rp{conditionRWA} is true,
it is legitimate to neglect the self correlations
$\av{\hat{c}_{w,T}^{(in)}(t)\ \hat{c}_{w,T}^{(in)}(t')}$ and $\av{\hat{c}_{w,T}^{(in)}{}\da(t)\ \hat{c}_{w,T}^{(in)}{}\da(t')}$. In fact, if we analyze the rotated input noise operators we see that the corresponding self correlation
$\av{\hat{c}_{w,T}^{\prime\,(in)}(t)\ \hat{c}_{w,T}^{\prime\,(in)}(t')}=\av{\hat{c}_{w,T}^{(in)}(t)\ \hat{c}_{w,T}^{(in)}(t')}\ e^{i\,\Delta_{w,\xi}(t+t')}$
exhibits a fast oscillating term
which makes its contribution negligible. In other terms, the microwave input noise operator of the transmitter can be safely approximated with thermal noise
with $\bar{n}_{w,T}\al{fb}$ thermal excitations.

\section{ -- The output fields in Fourier space}\label{App2}

The quantum Langevin equations~\rp{QLEs} can be easily solved in Fourier space. Specifically, given a generic operator $\hat x(t)$, one introduces the corresponding operator in Fourier space as $x(\omega)=\frac{1}{\sqrt{2\,\pi}}\int_{-\infty}^\infty\ \dd t\ \e^{i\,\omega\,t}\ \hat x(t)$, and its hermitian conjugate $\pq{x(\omega)}\da\equiv x\da(-\omega)$. In this way
one can transform Eq.~\rp{QLEs} and obtain the quantum Langevin equations in Fourier space
\begin{eqnarray}
0&=&
\ppt{i\,\omega-\kappa_{w,\xi}}\ c_{w,\xi}(\omega)-i\,G_{w,\xi}\ b_\xi(\omega)
+\mu_\xi\
c_{o,\xi}^{ \dagger}(\omega)
+\sqrt{2\ \kappa_{w,\xi}}\ c_{w,\xi}^{(in)}(\omega)\ ,
\nn\\
0&=&
-\ppt{i\,\omega+\kappa_{o,\xi}}\ c_{o,\xi}(-\omega)-i\,G_{o,\xi}\ b_\xi^{ \dagger}(-\omega)+\sqrt{2\ \kappa_{o,\xi}}\ c_{o,\xi}^{(in)}(-\omega)\ ,
\nn\\
0&=&
\ppt{i\,\omega-\gamma_{M,\xi}}\ b_\xi(\omega)-i\,G_{o,\xi}\ c_{o,\xi}^{\dagger}(\omega)-i\,G_{w,\xi}^{*}\ c_{w,\xi}(\omega)
\nn\\&&\hspace{0.5cm}
+\sqrt{2\ \gamma_{M,\xi}}\ b_{\xi}\al{in}(\omega)\ ,
\end{eqnarray}
where the input noise operators describe thermal noise with $\av{b_\xi\al{in}(\omega)\ b_\xi\al{in}{}\da(\omega')}=(1+\bar n_{M,\xi}\al{th})\delta(\omega+\omega')$, $\av{c_{i,\xi}\al{in}(\omega)\ c_{i,\xi}\al{in}{}\da(\omega')}=(1+\bar n_{i,\xi}\al{th})\delta(\omega+\omega')$ and $\av{c_{w,T}\al{in}(\omega)\ c_{w,T}\al{in}{}\da(\omega')}=(1+\bar n_{w,T}\al{fb})\delta(\omega+\omega')$.
These equations can be solved and, together with the standard input output relations $c_{i,\xi}\al{out,j}(\omega)=\sqrt{2\,\kappa_{i,\xi}\al{j}}\ c_{i,\xi}(\omega)-c_{i,\xi}\al{in,j}(\omega)$, can be used to  determine the expressions for the output field operators reported in Eqs.~\rp{Final output field (o,1)} and \rp{Final output field (w,1)}, where the coefficients are explicitly given by
\begin{eqnarray}
\mathcal{A}_\xi(\omega)&= &\chi_\xi(\omega)
\lpq{
\frac{i \vert G_{o,\xi}\vert^2G_{w,\xi}  \sqrt{2\gamma_{M,\xi}}}{(\kappa_{o,\xi}-i\omega)(\kappa_{w,\xi}-i\omega)(\gamma_{M,\xi}-i\omega)^2}\psi_\xi(\omega)
}\nn\\&&\rpq{
-\frac{iG_{o,\xi}\sqrt{2\gamma_{M,\xi}}}{(\kappa_{o,\xi}-i\omega)(\gamma_{M,\xi}-i\omega)}
}\ ,
\nn\\\label{B}
\mathcal{B}_\xi(\omega)&= &\frac{G_{o,\xi}G_{w,\xi} \sqrt{2\kappa_{w,\xi}\al{1}}}{(\kappa_{o,\xi}-i\omega)(\kappa_{w,\xi}-i\omega)(\gamma_{M,\xi}-i\omega)}\chi_\xi(\omega)\psi_\xi(\omega)\ ,
\\
\mathcal{C}_\xi(\omega)&= &\frac{G_{o,\xi}G_{w,\xi} \sqrt{2\kappa_{w,\xi}\al{2}}}{(\kappa_{o,\xi}-i\omega)(\kappa_{w,\xi}-i\omega)(\gamma_{M,\xi}-i\omega)}\chi_\xi(\omega)\psi_\xi(\omega)\ ,
\nn\\
\mathcal{D}_\xi(\omega) &= & \frac{\sqrt{2\kappa_{o,\xi}\al{1}}}{(\kappa_{o,\xi}-i\omega)}\chi_\xi(\omega)-1\ ,
\nn\\
\mathcal{E}_\xi(\omega)&= &  \frac{\sqrt{2\kappa_{o,\xi}\al{2}}}{(\kappa_{o,\xi}-i\omega)}\chi_\xi(\omega)\ ,
\nn
\end{eqnarray}
and
\begin{eqnarray}
\mathcal{A^{\prime}}_\xi(\omega)&= & \sqrt{2\kappa_{w,\xi}\al{1}}
\pq{
\frac{\mathcal{A}_\xi^{*}(-\omega)}{\sqrt{2\kappa_{o,\xi}\al{1}}}\Phi_\xi(\omega)
-\frac{iG_{w,\xi}\sqrt{2\gamma_{M,\xi}}}{(\kappa_{w,\xi}-i\omega)(\gamma_{M,\xi}-i\omega)}
}\psi_\xi(\omega)\ ,
\nn\\
\mathcal{B^{\prime}}_\xi(\omega)&= & \sqrt{2\kappa_{w,\xi}\al{1}}
\pq{
\frac{\mathcal{B}_\xi^{*}(-\omega)}{\sqrt{2\kappa_{o,\xi}\al{1}}}\Phi_\xi(\omega)
+\frac{\sqrt{2\kappa_{w,\xi}\al{1}}}{(\kappa_{w,\xi}-i\omega)}
}\psi_\xi(\omega)-1\ ,
\nn\\
\mathcal{C^{\prime}}_\xi(\omega)&= & \sqrt{2\kappa_{w,\xi}\al{1}}
\pq{
\frac{\mathcal{C}_\xi^{*}(-\omega)}{\sqrt{2\kappa_{o,\xi}\al{1}}}\Phi_\xi(\omega)
+\frac{\sqrt{2\kappa_{w,\xi}\al{2}}}{(\kappa_{w,\xi}-i\omega)}
}\psi_\xi(\omega)\ ,
\nn\\
\mathcal{D^{\prime}}_\xi(\omega)&= & \sqrt{\frac{\kappa_{w,\xi}\al{1}}{\kappa_{o,\xi}\al{1}}}
\pq{
\mathcal{D}_\xi^{*}(-\omega)+1
} \Phi_\xi(\omega)\psi_\xi(\omega)\ ,
\nn\\
\mathcal{E^{\prime}}_\xi(\omega)&= & \sqrt{\frac{\kappa_{w,\xi}\al{1}}{\kappa_{o,\xi}\al{1}}}\mathcal{E}_\xi^{*}(-\omega) \Phi_\xi(\omega)\psi_\xi(\omega)\ ,
\nn
\end{eqnarray}
with $\psi_\xi(\omega)$, $\chi_\xi(\omega)$ and $\Phi_\xi(\omega)$ defined as follows
\begin{eqnarray}
\psi_\xi(\omega)&= &
\pq{
1+\frac{\vert G_{w,\xi}\vert ^2}{(\kappa_{w,\xi}-i\omega)(\gamma_{M,\xi}-i\omega)}
}^{-1}\ ,
\nn\\
\chi_\xi(\omega)&= & \sqrt{2\kappa_{o,\xi}\al{1}}
\lpq{
1-\frac{\vert G_{o,\xi}\vert^2}{(\kappa_{o,\xi}-i\omega)(\gamma_{M,\xi}-i\omega)}
}\nn\\&&\rpq{
-\frac{G_{o,\xi}G_{w,\xi}}{(\kappa_{o,\xi}-i\omega)(\gamma_{M,\xi}-i\omega)}\psi_\xi(\omega)\Phi_\xi^{*}(-\omega)
}\ ,
\nn\\
\Phi_\xi (\omega)&=&
\frac{\mu_\xi}{(\kappa_{w,\xi}-i\omega)}
-\frac{G_{o,\xi}G_{w,\xi}}{(\kappa_{w,\xi}-i\omega)(\gamma_{M,\xi}-i\omega)}\ ,
\nn
\end{eqnarray}
where $\mu_\xi$ is defined in Eq.~\rp{muxi}.

\section{ -- The signal to noise ratio for microwave QI}
\label{App3}

Here we briefly discuss the derivation of the signal to noise ratio in Eq.~\rp{QI SNR}. In details, we consider the expression for the signal to noise ratio (see Sec.~\ref{Sec.MQI})
\begin{eqnarray}\label{SNR.App}
SNR_{QI}\al{M}=4\, M\ppt{\frac{N\bigl|_{H_1}-N\bigl|_{H_0}}{
\sigma\bigl|_{H_1}+\sigma\bigl|_{H_0}
}}^2\ .
\end{eqnarray}
The average numbers in the numerator can be expressed as
\begin{eqnarray}
N\bigl|_{H_j}=\av{N_+}_{H_j}-\av{N_-}_{H_j}
\end{eqnarray}
with
\begin{eqnarray}
\av{N_\pm}_{H_j}=\int\,\dd\omega'\ \av{d_\pm\da(\omega)\ d_\pm(\omega')}_{H_j}\ ,
\end{eqnarray}
where $d_\pm$ are the operators for the modes after the beam splitter which mixes the idler and the reflected, converted and phase-conjugated signal,
\begin{eqnarray}
d_\pm=\frac{c_{o,R}\al{out,1}\pm c_{o,T}\al{out,1}}{\sqrt{2}}\ ,
\end{eqnarray}
so that
\begin{eqnarray}
N\bigl|_{H_j}&=&
\int\,\dd\omega'\
\av{c_{o,R}\al{out,1}{}\da(\omega)\ c_{o,T}\al{out,1}(\omega') + c_{o,T}\al{out,1}{}\da(\omega)\ c_{o,R}\al{out,1}(\omega')}\Bigl|_{H_j}
\nn\\&=&
\int\,\dd\omega'\
\BBB_R^*(-\omega)\ \av{c_{w,R}\al{in,1}(\omega)\ c_{o,T}\al{out,1}(\omega') }_{H_j}
\nn\\&&+
\int\,\dd\omega'\
\BBB_R(-\omega) \av{c_{o,T}\al{out,1}{}\da(\omega)\ c_{w,R}\al{in,1}{}\da(\omega')}_{H_j}
\ .
\end{eqnarray}
Since, according to the QI protocol, as discussed in Sec.~\ref{Sec.MQI},
\begin{eqnarray}
c_{w,R}\al{in,1}=
\lpg{
\mmat{cc}{
c_B&\ {\rm for}\ j=0\\
\sqrt{t}\ c_{w,T}\al{out,1}+\sqrt{1-t}\ c_B&\ {\rm for}\ j=1
}
}
\end{eqnarray}
one finds
\begin{eqnarray}\label{N}
N\bigl|_{H_j}&=&
\lpg{
\mmat{cc}{
0&\ {\rm for}\ j=0\\
\sqrt{t}\ \pq{ \BBB_R^*(-\omega)\ m_T(\omega) +\BBB_R(-\omega)\ m_T(\omega)^* } &\ {\rm for}\ j=1\ ,
}
}
\end{eqnarray}
where $m_T(\omega)$ is introduced in Eq.~\rp{m}.
Moreover, the variances in the denominator of Eq.~\rp{SNR.App} can be expressed
as
\begin{eqnarray}\label{sigma}
\sigma\Bigl|_{H_j}&=&\lpg{  \phantom{\pq{\int}^2}\hspace{-0.9cm}
\av{N_+}_{H_j}\ppt{\av{N_+}_{H_j}+1}+
\av{N_-}_{H_j}\ppt{\av{N_-}_{H_j}+1}
}\nn\\&&\hspace{-1.5cm}
-
\frac{1}{2}\pq{
\int\,\dd\omega'\
\av{c_{o,R}\al{out,1}{}\da(\omega)\ c_{o,R}\al{out,1}(\omega')}_{H_j}-
\av{c_{o,T}\al{out,1}{}\da(\omega)\ c_{o,T}\al{out,1}(\omega')}
}^2
\nn\\&&\hspace{-1.5cm}\rpg{
+2\,\pq{
\int\,\dd\omega'\
Im\,
\av{c_{o,R}\al{out,1}{}\da(\omega)\ c_{o,T}\al{out,1}(\omega')}
}^2
}^{1/2}\ .
\end{eqnarray}
In particular, in the limit of large $N_B=\av{c_B\da\ c_B}$, one can retain only the terms proportional to $\sqrt{N_B}$ and neglect the others, so that one finds the approximation
\begin{eqnarray}
\sigma\Bigl|_{H_1}\simeq\sigma\Bigl|_{H_0}\simeq \abs{\BBB_R(-\omega)}\ \sqrt{N_B\pq{1+2\,n_{o,T}(-\omega)}}\ ,
\end{eqnarray}
with $n_{o,T}$ defined in Eq.~\rp{n}.
Finally, using Eqs.~\rp{N} and \rp{sigma} in Eq.~\rp{SNR.App} one finds Eq.~\rp{QI SNR}.

\end{document}